\documentclass[a4paper,11pt]{article}
\usepackage[pdftex]{graphicx}
\usepackage{amsmath,amssymb,color,verbatim,fullpage}
\usepackage[affil-it]{authblk}

\usepackage[numbers,sort&compress]{natbib}
\usepackage{comment,caption}
\usepackage[usenames,dvipsnames]{xcolor}
\usepackage[normalem]{ulem}

\begin{document}
\title{Time-Dependent Flow in Arrested States ---\\ Transient Behaviour}
\author{K J Mutch$^{a *}$, M Laurati$^{a}$, C P Amann$^{b}$, M Fuchs$^{b}$ and S U Egelhaaf$^{a}$ \\ {\footnotesize{(a) Condensed Matter Physics Laboratory, Heinrich-Heine University, \\40225 D\"usseldorf, Germany}} \\{\footnotesize{(b) Fachbereich Physik, University of Konstanz, 78457 Konstanz, Germany}} \\{\footnotesize{* kevin.mutch@uni-duesseldorf.de}}}
\date{}

\maketitle
\abstract{
The transient behaviour of highly concentrated colloidal liquids and dynamically arrested states (glasses) under time-dependent shear is reviewed. This includes both theoretical and experimental studies and comprises the macroscopic rheological behaviour as well as changes in the structure and dynamics on a microscopic individual-particle level. The microscopic and macroscopic levels of the systems are linked by a comprehensive theoretical framework which is exploited to quantitatively describe these systems while they are subjected to an arbitrary flow history. Within this framework, theoretical predictions are compared to experimental data, which were gathered by rheology and confocal microscopy experiments, and display consistent results. Particular emphasis is given to (i) switch-on of shear flow during which the system can liquify, (ii) switch-off of shear flow which might still leave residual stresses in the system, and (iii) large amplitude oscillatory shearing. The competition between timescales and the dependence on flow history leads to novel features in both the rheological response and the microscopic structure and dynamics.
}
\section{Introduction}
\label{sec:intro}

The ubiquity of colloidal dispersions, coupled with their ability to act as suitable models for atomic and molecular systems~\cite{Pusey_91}, make them both academically and commercially very relevant. One of the primary advantages of studying colloidal systems is their size: they are often large enough to be seen with a microscope and slow enough to be followed. Moreover, the typical shear modulus of a colloidal dispersion is orders of magnitude lower than equivalent atomic and molecular systems~\cite{Poon_00,McLeish_00}, rendering the flow properties of such systems experimentally accessible. This has relevance to industrial processes, where materials are often dispersed as colloidal particles in order to pour or transport them, dispersion paints being just one familiar example. It is thus not surprising that the rheology of dispersions has been the focus of longstanding research and a number of approaches have been developed in this field, some of which were recently reviewed~\cite{brader2010b}.

For the case of hard sphere colloids, the thermodynamic control parameter is the volume fraction of particles, $\phi$. In a monodisperse system, the colloids exhibit a disordered fluid phase until freezing at $\phi=0.494$, whereby a first-order phase transition to a crystal phase of $\phi=0.545$ occurs~\cite{Hoover_68,Pusey_86,Gasser_09}. An increase in the polydispersity of the colloidal dispersion leads to a frustration of the crystalline ordering and suppression of crystallisation for polydispersities greater than $\sim 7$\%~\cite{Pusey_09}. Instead, a non-equilibrium glassy state is observed for $\phi\geq \phi_{\mathrm{g}} \approx 0.58$~\cite{Woodcock_81,Pusey_86}, at which point the dynamics become arrested.

Such concentrated colloidal systems can be conveniently characterised  by dynamic light scattering (DLS) and confocal microscopy experiments. Observation of the intermediate scattering function (ISF) from DLS reveals a non-decaying component upon increasing $\phi$ beyond the glass transition, suggesting a dynamical slowing down~\cite{Pusey_87,vanMegen_91}. The mode-coupling theory (MCT)~\cite{Goetze_92,Goetze} has been shown to accurately describe these experimentally determined ISF~\cite{vanMegen_93,vanMegen_94}. Mean-squared displacements (MSD) can be derived from the ISF and can also be measured by confocal microscopy; the slowing down in the dynamics as $\phi$ increases is evidenced by an emerging plateau in the MSD, through which a localisation length related to particle caging can be determined. In the concentrated disordered fluid state, particles are temporarily caged by their neighbours but can in time escape these cages and display diffusion at long times. As the concentration increases towards and above the glass transition, the cages tighten and the particles become immobilised; their only motion is local movement within the cages. The restricted mobility has consequences for their response to mechanical stress, such as reversible elastic deformation and resistance to flow at small applied stresses followed by yielding and irreversible deformation at large applied stresses. 

The low yield stress of colloidal glasses, i.e.~their mechanical weakness (or softness), imparts interesting and varied behaviour, particularly under conditions of flow. A long observed phenomenon in the sheared steady state is that of shear thinning, i.e.~the decrease of the shear viscosity with increasing rate of shear~\cite{Krieger_72,Krieger_76,Ackerson_91,Pham_06,Pham_08,Sentjabrskaja_13}. In order to understand such macroscopic behaviour, it is necessary to draw a link to the microstructure of the system~\cite{Batchelor_77}, which can be achieved through modern scattering and imaging techniques. For example, the application of light scattering echo~\cite{Petekidis_02A,Petekidis_02,Petekidis_03,Pham_04} and more recently confocal microscopy~\cite{Besseling_07, Mutch_13} has allowed the origins of shear thinning to be better understood. From microscopy experiments, it was seen that the inverse relaxation time of a sheared glass was closely linked to the local rather than global shear rate. The converse case of shear thickening can also arise at high shear rates due to cluster formation induced by increased hydrodynamic interactions~\cite{DHaene_93,Bender_96,Maranzano_02,Foss_00,Brady_01}. 

These studies of colloidal systems under flow, i.e.~in the steady state of shear, effectively require ``pre-yielding'' and  a certain strain to have been reached. In order to achieve this however, first the linear regime must be traversed, in which the stress $\sigma$ is proportional to strain $\gamma$, followed by the transient regime. It is in this transient regime of intermediate strains that some particularly interesting phenomena can occur. The transport coefficients measured in quiescent equilibrium or linear response are known to differ greatly from those in steady shear states and hence the microscopic properties of a system exposed to a sudden change in externally applied shear are of particular interest. Moreover, the transient regime encompasses the shear melting of colloidal glasses and could potentially provide greater understanding of glassy states at rest.

First insight to the transient response can be gleaned by measuring the evolution of the stress, $\sigma$, under conditions of constant shear rate, $\dot{\gamma}$. For certain colloidal systems, the stress evolves non-monotonically~\cite{Nguyen_83,Persello_94,Derec_03,Mahaut_08}. Upon start-up of shear, the stress grows almost linearly with time before reaching a maximum value and subsequently decaying to the steady state asymptotic value. The intermediate maximum is referred to as the \emph{stress overshoot} and has also been observed in other materials displaying a yield stress~\cite{Khan_88,Islam_04,Mohraz_05,Becu_06,Akcora_09,Carrier_09,Divoux_11}. 
The ability of a system to yield is not a pre-requisite to display an overshoot however, as systems of entangled polymers and wormlike micelles also exhibit such behaviour~\cite{Berret_97,Lerouge_98,Soltero_99,Osaki_00,Islam_01,Decruppe_01,Boukany_10,Ganapathy_08,Ravindranath_08,Tapadia_04}; it has been proposed that these systems display yielding like behaviour due to the disentangling of polymer/micelle chains~\cite{Tapadia_04,Tapadia_06}. Conversely, a microscopic understanding of yielding and the presence of a stress overshoot in dense colloidal systems is not so clear. Hence, a discussion of the rheology, structure and dynamics of colloidal suspensions during the transient regimes of both start-up and cessation of flow, along with the periodic switching on and off of shear in the form of large-amplitude oscillatory shear (LAOS), will form the focus of this review.

\section{Theory}
\label{sec:theory}

\subsection{Overview of theoretical approaches}

Theoretical treatment of the rheology of dense colloidal suspensions generally falls into one of three main categories, each of which will be briefly summarised. 

Firstly, phenomenological approaches are based on kinetic rules or dynamical equations for coarse-grained fields. These approaches traditionally start from continuum hydrodynamics by formulating constitutive equations for the stress-strain relations and closing the Navier-Stokes equations~\cite{larson}. This led to, e.g., the `principle of material objectivity', which constrains constitutive equations to be invariant under arbitrary (including time-dependent) solid body rotations. Nevertheless, a detailed microscopic picture of local structural deformations under flow has not been provided.

In the mesoscopic regime, other phenomenological approaches like shear transformation zone (STZ) theory have been applied to linear dispersion moduli~\cite{Langer2011}. STZ considers kinetic rate equations for density fields of flow defects which, when present in small numbers, could be observed in colloidal glasses using confocal microscopy~\cite{Schall_07}. The existence of flow defects at densities below the glass transition is, however, still unclear. Soft glassy rheology (SGR) is another phenomenological approach, whereby a set of elastic elements with distributed yield thresholds are considered~\cite{SGR}. They undergo activated dynamics governed by a noise temperature, which differs from the thermodynamic one. The microscopic basis of this model for colloids has not yet been established, and new results indicate that SGR is more adapted to jamming or athermal systems such as granular materials~\cite{Siebenbuerger_12,Ikeda_12}. They exhibit different shear-thinning behaviour than dense colloidal dispersions, which is discussed in the present review.

Other microscopic approaches start from the underlying stochastic processes, where the colloids display Brownian motion in an incompressible  solvent. A multi-particle stochastic equation (Smoluchowski equation) captures the dynamics and rheology of colloidal dispersions for times longer than the diffusion time $\tau_0=a^2/D_0$, i.e.~the time it takes an isolated colloid to diffuse its own radius $a$~\cite{dhont}. Here, $D_0=k_BT/(6\pi\eta_s a)$ is the Stokes-Einstein-Sutherland diffusion coefficient, which can be determined from the solvent viscosity $\eta_s$.

The second class of approaches specifically addresses the low density limit, where consideration of the two-particle distribution function suffices~\cite{Foss_00,Brady_01,brady,Russel}. The resulting theory for the relative density distributions, stress and flow fields can be solved rigorously, providing insight into shear-thinning and shear-thickening phenomena at low densities. The complete inclusion of hydrodynamic interactions has been achieved~\cite{brady}; these interactions propagate through the solvent instantaneously, causing both lubrication forces in the near-field and long-ranged force-velocity correlations in the far-field. For large flow rates, the distortion of the probability of particles to approach each other varies on two very different length scales: for large separations, flow randomizes the correlations between the particle-pair, whilst for small separations, diffusion and flow compete in a narrow boundary layer at contact. For flows at a steady flow rate $\dot\gamma$, the competition is characterized by the bare P\'eclet number $Pe_0=\dot\gamma \tau_0=\dot\gamma a^2/D_0$, i.e.~the ratio between the rates of external flow and internal diffusion. Asymptotically, the boundary layer obtains a width ${\cal O}({\rm Pe}_0^{-1})$ and height that increases with $Pe_0$. In the strongly varying local density field, local stress fields arise and determine the macroscopic rheological properties~\cite{Russel,brady}. For colloidal hard spheres, the density increase is measured by the contact value of the pair-correlation function. This aspect is the basis for the extension to higher packing fractions~\cite{brady}. The binary-correlation results obtained at low densities are scaled  into the semi-dilute regime using the packing fraction dependent contact value. This scaling has also been suggested to hold close to random-close packing, where the contact value diverges~\cite{Brady_93}. Because this scaling focuses on the local packing, which is connected to motion characterised by the short-time self-diffusivity, the slow structural relaxation is neglected. The latter are connected to the long-time relaxation in the supercooled fluid. The approach~\cite{brady} therefore does not apply to the region around the glass transition, where cooperative structural rearrangements reaching beyond the boundary layer dominate.

The third and final group of theories are those which directly address concentrated dispersions and their density fluctuations through a microscopic approach. For example, time dependent density functional theories (DDFT) have been intensively pursued~\cite{brader2011}. Whilst some DDFT approaches contain an instability leading to glassy arrest, the incorporation of structural memory and slow dynamics remains unresolved. If the system at low enough density escapes the DDFT arrest-phenomenon, its long-time diffusivity returns to the value at infinite dilution. At long times, the supercooled fluid has no memory of the incipient glassy arrest in DDFT~\cite{Archer}, which is clearly at odds with dynamic light scattering observations~\cite{vanMegen1998}. Another route to the nonlinear rheology of dense dispersions builds on the mode coupling theory (MCT) of the glass transition~\cite{Goetze_92,Goetze} and extends it to nonequilibrium phenomena via the integration through transients (ITT) approach~\cite{Kawasaki,EvansITT}. The MCT-ITT approach is focused on the slow structural relaxation phenomena under weak flows which include shear-thinning as the main effect in complex fluids. Recent developments in the generalisation of MCT-ITT to incompressible and homogeneous but otherwise arbitrary weak flows in dense dispersions~\cite{brader2008,brader2012}, where weak entails $Pe_0\ll1$, are reviewed in the following section.

\subsection{Constitutive relations of microscopic MCT-ITT}

The MCT-ITT approach was developed in order to describe steadily sheared dispersions where internal structural relaxation competes with an external driving force~\cite{fuchs02,fuchs09}. It was recently generalized to arbitrary incompressible and homogeneous weak flows~\cite{brader2008,brader2012}, in order to incorporate transient phenomena and general flow geometries. Shear thinning remains the dominating phenomenon captured in the MCT-ITT equations. It results from the flow advection of density fluctuations, which are captured in wavevector space. Consideration of spatial Fourier-modes has proven to be an efficient method to describe structural relaxation, presumably because glassy arrest results from a nonlinear scattering process where plane wave density fluctuations at wavevectors $\bf p$ and $\bf k$ scatter and excite a fluctuation at $\bf q=p+k$.  The wavevectors become time-dependent due to flow advection because the affine particle motion under flow tilts or distorts the plane wave modes. Describing the homogeneous flow with a velocity gradient tensor $\boldsymbol \kappa(t)$, the advected wavevector 
\begin{eqnarray}
{\bf k}(t,t')={\bf k}\cdot e_+^{\int_{t'}^{t}\!ds\,\boldsymbol \kappa (s)}
\label{advection2}
\end{eqnarray}
enters the theory. Here, the time dependence of the flow between the times $t'$ and $t$ is given by the (time-ordered) exponential that also describes the deformation in  real space:
 \begin{eqnarray}\label{deformation}
{\bf E}(t,t')=\frac{\partial {\bf r}(t)}{\partial {\bf r}(t')} = e_+^{\int_{t'}^{t}\!ds\,\boldsymbol \kappa (s)}
\end{eqnarray}
The tensor $\bf E$ is the deformation gradient tensor~\cite{larson}. The principle of material objectivity, which requires that the nonlinear rheology be independent of solid-body rotations of the sample, is verified in MCT-ITT because at each appearance of $\bf E$ it can be combined into the finger tensor ${\bf B}(t,t')={\bf E}(t,t')\cdot {\bf E}^T(t,t')$, which is manifestly invariant~\cite{brader2008,brader2012}.

In order to establish a constitutive relation which links stress and strain, in MCT-ITT first a formally exact generalized Green-Kubo relation for the macroscopic stress tensor is derived. It can be considered a formally exact constitutive equation under homogeneous flow
\begin{eqnarray}
\boldsymbol \sigma(t) = \frac{1}{V}\int_{-\infty}^{t}\!\!\!\!\! dt_1\,
\langle \boldsymbol \kappa(t_1) \!:\!\hat{\boldsymbol \sigma} \,e_-^{\int_{t_1}^{t}ds\,\Omega^{\dagger}(s)} \hat{ \boldsymbol\sigma}\rangle
\label{stress}
\end{eqnarray}
Here, $\hat{\boldsymbol \sigma}$ is the fluctuating stress tensor. The averaged stress tensor is a nonlinear functional of the velocity gradient tensor as $\boldsymbol \kappa(t)$ appears in the adjoint Smoluchowski operator  $\Omega^{\dagger}$~ \cite{brader2008,brader2012}. 

Building on the insights gained through MCT into the cage-effect in supercooled liquids \cite{Goetze}, approximations are  formulated which enable evaluation of the formally exact Green-Kubo relation. The first approximation expresses the stress fluctuations in Eq.~\eqref{stress} in terms of pair density fluctuations. Stresses are supposed to arise from density fluctuations and to become slow because of the slow structural rearrangements which are captured in the ensemble averaged transient density correlators
\begin{eqnarray}
\Phi_{{\bf k} (t,t')}(t,t') = \frac{\langle \rho^{*}_{{\bf k} (t,t')} e_-^{\int_{t'}^t ds\,\Omega^{\dagger}(s)} \rho_{\bf k}^{}\rangle}{N S_{k(t,t')}} \label{correlator_def}
\end{eqnarray}
The forward-advected wavevector, Eq.~\eqref{advection2}, appears naturally in Eq.~\eqref{correlator_def}. The correlator arises in the approximated stress, Eq.~\eqref{stress}, after a Wick-like decomposition of a four-density correlation function into the product of two density fluctuations that make up the transient correlator. The final result takes the form:
\begin{eqnarray}
\boldsymbol \sigma(t) &=& -\int_{-\infty}^{t} \!\!\!\!\!\!dt'\!\int\!\!\!
\frac{d{\bf k}}{32\pi^3} \left(\frac{\partial}{\partial t'}({\bf k}\!\cdot\!{\bf B}(t,t')\!\cdot\!{\bf k})\right)\; 
\frac{{\bf k}{\bf k}}{k}\!
\left(\!\frac{S'_k S'_{k(t,t')}}{k(t,t')S^2_k}\!\right)\Phi_{{\bf k}(t,t')}^2(t,t')
\label{nonlinear}
\end{eqnarray}    
The MCT equations of motion are fully specified by the equilibrium structure factor $S_q$ and a single initial time scale. It is often connected to the short time diffusion coefficient. Eq.~\eqref{nonlinear} thus predicts the nonlinear rheology from equilibrium structural correlations, and neglects hydrodynamic interactions; MCT presumes that they only shift the time scale and lead to a high-frequency viscosity $\eta_\infty$. The transient density correlator in Eq.~\eqref{nonlinear} depends only on the magnitude of the advected wavevector, rendering the MCT-ITT result dependent exclusively on the finger tensor and hence resulting in material objectivity~\cite{brader2012}. Its validity in the exact Eq.~\eqref{stress} was shown by a discussion of the transient correlation function \cite{brader2012}.

\subsection{Schematic MCT-ITT equations}

Structural correlations at the peak of the structure factor dominate the microscopic MCT-ITT equations. These are connected to the average particle separation via the inverse of the wave vector The spatially resolved equations for $\Phi_{\bf q}(t,t')$ can be simplified to a single equation for a local density correlator $\Phi(t,t')$, because the dynamics on all length scales are coupled strongly (factorization theorem of MCT~\cite{Goetze}). In the resulting schematic model \cite{brader2009,Amann_13}, the modulus $G(t,t')$, which determines the shear stress via causal temporal integration
\begin{equation}
\sigma(t) = \int_{-\infty}^{\,t}\!\!dt'\; \dot\gamma(t')\,G(t,t')
\label{non-tti}
\end{equation}
is expressed by the transient correlator as
\begin{equation}\label{eqn:gschem}
  G(t,t')=v_{\sigma}(t,t')\, \Phi^2(t,t')
\end{equation}
The transient (shear) stress-stress correlation function $G(t,t',[\dot\gamma])$ depends on the full flow history arising from the time-dependent shear rate $\dot\gamma(t)$. In general, it depends on the two times corresponding to the underlying fluctuations separately. The strain-dependent function $v_\sigma$ is an elastic coefficient that captures the coupling of stress to density fluctuations and was set to a constant in the original formulation of the model~\cite{brader2009}. In the microscopic theory, Eq.~\eqref{nonlinear}, $G(t,t')$ is given by an integration over wave vectors, including nontrivial weights that depend on time, since density fluctuations are advected by shear. To account for the dephasing of wavevector contributions, which render $G$ negative in the microscopic theory, a more refined model~\cite{Amann_13} allows to express the time-dependence of the prefactor $v_\sigma(t,t')$ through the accumulated strain $\gamma(t,t')=\int_{t'}^t\dot\gamma(s)\,ds$:
\begin{equation}
  v_{\sigma}(t,t')=v_{\sigma}^*\, \cdot\left( 1-\left(\frac{\gamma\left(t,t'\right)}{\gamma^{*}}\right)^2\right)\, \exp\left(-\left(\frac{\gamma\left(t,t'\right)}{\gamma^{**}}\right)^2\right).\label{eqn:ammanprefactor}
\end{equation} 
The direction of the strain does not matter due to symmetry, leading to an even function $v_\sigma=v_\sigma(\gamma(t,t')^2)$. The two material parameters $\gamma^\ast$ and $\gamma^{\ast\ast}$ denote characteristic strain values. If the accumulated strain exceeds $\gamma^\ast$, stress fluctuations become anticorrelated and the transient correlator in Eq.~\eqref{eqn:gschem} becomes negative, corresponding to the stress maximum. Beyond $\gamma^{\ast\ast}$ stress fluctuations have decorrelated and $G(t,t')$ vanishes, i.e. the steady state stress value has been reached. Eq.~\eqref{eqn:ammanprefactor} was justified by a comparison with fully microscopic MCT-ITT calculations in two dimensions \cite{Amann_13}. The single-mode density correlator (normalized to $\Phi(t,t)=1-\Gamma(t{-}t')$)  is obtained from an equation of motion simplified from the full MCT equations by again neglecting wave vector dependences.

\section{Start-up of shear: switch on}
\label{sec:swon}

\subsection{Rheology}
\label{sec:swon-rheo}

\begin{figure}[ht]
\begin{center}
\includegraphics[width=0.7\linewidth]{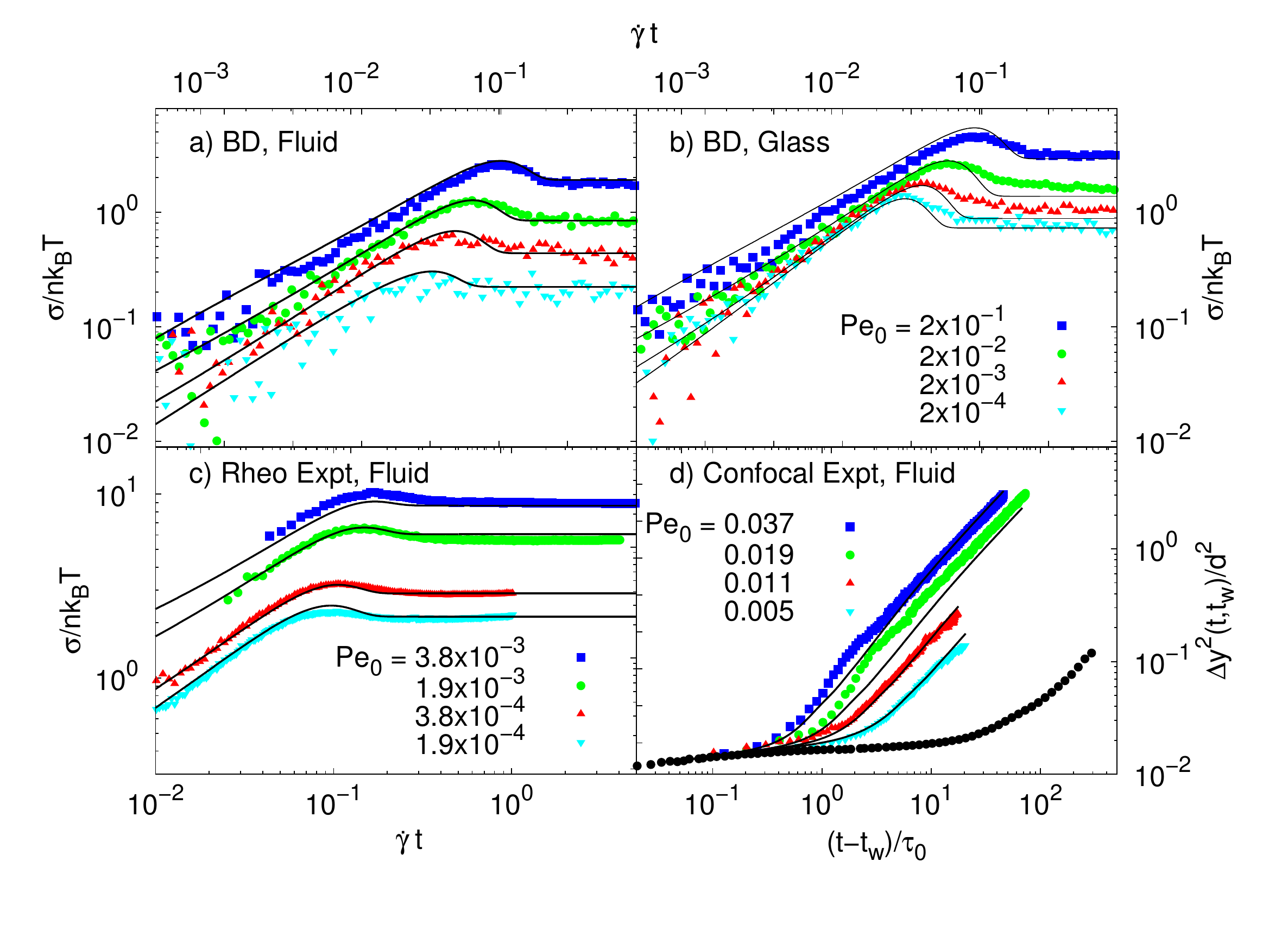}
\caption{(a,b) Stress vs. strain curves from Brownian dynamic simulations in two dimensions for various values of $Pe_0$, as labelled, and packing fractions $\phi=0.79$ (a) and $\phi=0.81$ (b) \cite{Amann_13}. (c) Stress vs. strain curves from PMMA particles of $a=267$~nm and $\phi=0.56$. (d) Switch-on MSD in the vorticity direction for waiting time $t_w=0$~s for PMMA particles with $a=780$~nm, $\phi=0.56$ and $Pe_0$ as indicated \cite{Laurati_12}. In all panels, solid black lines are fits from the $F_{12}^{(\dot\gamma)}$ model. \label{fig:sw-on}}
\end{center}
\end{figure}

During a start-up shear experiment of fixed shear rate $\dot{\gamma}$ the evolution of the shear stress $\sigma$ can be followed with the progression of time, or equivalently strain $\gamma=\dot{\gamma}t$. Brownian dynamics simulations (Fig.~\ref{fig:sw-on}(a,b)~\cite{Amann_13}) and rheological experiments (Fig.~\ref{fig:sw-on}(c)~\cite{Laurati_12}) display three distinct regimes: the linear, the transient and the steady state. In the linear regime, the stress grows almost linearly with increasing strain in an elastic-like fashion before, during the transient regime, reaching a maximum $\sigma_{max}$ at an intermediate strain $\gamma_{max}$. Finally, there is a stress decay to an asymptotic value in the steady state, $\sigma_{\mathrm{ss}}({\dot\gamma})$, where the system exhibits a fluidlike response. 

The stress overshoots observed in experiments of hard-sphere-like poly-methyl\-meth\-acrylate (PMMA) particles, molecular dynamic (MD) simulations and Brownian dynamic (BD) simulations, were compared by Laurati et al.~\cite{Laurati_12}. It was noted that both the relative magnitude of the overshoot and the strain at which it occurs increase with increasing $Pe_0$. The diminishing ability of Brownian motion to fully relax the distorted structures as $Pe_0$ rises is thought to be the reason for this change. Therefore, systems exposed to higher shear rates are able to store more stress before yielding. Upon increasing the volume fraction, the stress overshoot becomes less prominent as the free volume around a particle decreases~\cite{Koumakis_12a}. Upon approaching random close packing, where the system is fully jammed, the overshoot essentially disappears due to the inability of particle cages to deform and hence store stress. 

An alternative experimental system of colloids with a polystyrene core and a temperature sensitive shell of crosslinked poly(N-isopropylacrylamide) (PNIPAM), which shrinks upon increasing temperature, have also been studied by rheology~\cite{Crassous_08,Siebenbuerger_09,Siebenbuerger_12,Amann_13}. A similar increase in the magnitude of the stress overshoot with increasing $Pe_0$ was observed~\cite{Amann_13}, in agreement with the behaviour of PMMA systems~\cite{Laurati_12,Koumakis_12a}. Upon increasing $\phi$ however, the softer PNIPAM particles display an overshoot up to very high volume fractions, unlike the PMMA system. The origin of this difference is unclear but could be explained by particle softness; the PNIPAM shells have the ability to compress and hence the system can still store stress through an overall shell deformation~\cite{Koumakis_12a}. Experimental data from both the PMMA and PNIPAM systems as well as simulation data can be well described by MCT~\cite{Laurati_12,Amann_13} (lines in Fig.~\ref{fig:sw-on}).

\subsection{Microscopic Structure}
\label{sec:swon-str}

In order to further understand the mechanisms underlying stress storage and yielding, it is necessary to study the microscopic properties on the particle level. Whilst this has traditionally been possible through computer simulations, such as molecular dynamics (MD)~\cite{Varnik_04,Zausch_08} or Brownian dynamics (BD)~\cite{Koumakis_12,Amann_13}, it required recent developments in confocal microscopy to provide experimental access to such quantities~\cite{vanBlaaderen_95,Weeks_99,Dinsmore_01}. Through the use of either a purpose built shear cell~\cite{Derks_04,Smith_07,Besseling_07,Chen_10,Cheng_11}, or the coupling of a rheometer with a microscope~\cite{Ballesta_08,Besseling_09,Besseling_10,Dutta_13}, it is now possible to observe colloidal dispersions under shear. Use of tracking algorithms~\cite{Crocker_96,Jenkins_08} allow the locations and trajectories of particles to be determined and hence give access to local dynamical and structural information. 

The relation between the stress-overshoot and yielding can be clarified by considering the local structure under shear. Experimentally, this requires the measurement of 3-dimensional image stacks, which implies moving either the sample or the objective in order to observe adjacent focal planes. For this reason, the time resolution is lower compared to the observation of a single 2D plane; determining the structure of suspensions under shear hence requires either very low shear rates~\cite{Schall_07} or very high scanning rates~\cite{Besseling_07,Chen_10}. Using  an acousto-optic deflector to control the scanning of the laser beam of the confocal microscope~\cite{Semwogerere_05}, it is possible to quickly image colloidal suspensions under shear and directly observe the structural deformation which takes place in the transient regime.

\begin{figure}[htb]
\begin{center}
\includegraphics[width=0.68\linewidth]{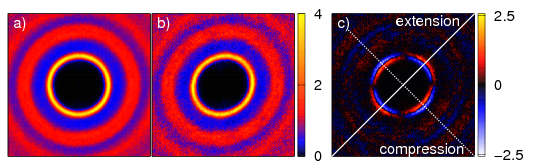}
\caption{2-dimensional $g(r)$ in the shear-gradient plane for PMMA particles of $a=780$~nm and $\phi=0.56$ obtained in (a) the quiescent, unsheared state, (b) the steady state of shear with $Pe_0$=0.0074 and (c) their difference, i.e.~the unsheared case subtracted from the sheared case. \label{fig:str}}
\end{center}
\end{figure}

Fig.~\ref{fig:str} shows the projection of the pair-distribution function, $g(r)$, in the quiescent, unsheared state and in the steady state of shear in the shear-gradient plane, along with the difference of the two \cite{Koumakis_long}. These results are similar to those observed by BD and MD simulations, where the structure was also directly compared to the stress evolution under shear~\cite{Zausch_09,Koumakis_12}. The results show that at very low strains ($\gamma\approx1\%$), the structure displays only a slight distortion but this grows with increasing strain, showing a shift of the first peak along the extension axis, corresponding to a reduced probability of finding particles at contact in this direction, but an increased probability along the compression axis due to the squeezing of the particle cage. At strains corresponding to the peak stress, the maximum structural anisotropy is observed before displaying more diffuse lobes along the extension axis with further increase of the strain~\cite{Zausch_09,Koumakis_12}. In the steady state, the constant distortion of the nearest neighbour cage is caused both by the interplay of increased particle escape along the extension axis and crowding along the compression axis, and the continuous balance between cage destruction and reformation. 

The value of the first maximum in $g(r)$ gives a measure of the probability of particles contacting and displays deviating behaviour along the compression and extension axes as strain increases. While the probability of contact increases on the compression axis due to the squeezing of the particle cage, the stretching of the cage in the perpendicular direction results in a lower probability of contact and hence a smaller value of $g(r)$. The deviation is most prominent near the stress maximum. The change in the value of $g(r)$ can be closely linked to the evolution of stresses in the system~\cite{Brady_93}, despite these being transient states~\cite{Koumakis_12}. This indicates a direct correlation of the structural anisotropy with the transient macroscopic stress during yielding~\cite{Koumakis_12, Koumakis_long}.

\subsection{Microscopic Dynamics}
\label{sec:swon-dyn}

The transient dynamics of colloidal dispersions under shear, as measured by confocal microscopy and in MD simulations, were first described by Zausch et al.~\cite{Zausch_08}. Experimentally, observation of a single plane of the sample through application of fast scanning confocal microscopy, allows particles to be followed with good time resolution and the MSDs to be calculated. Further development of this technique allowed a wider range of shear rates to be studied \cite{Laurati_12}.

Fig.~\ref{fig:sw-on}(d) shows the MSDs in the vorticity direction for varying shear rates as measured in the quiescent state, the steady-state and the transient state starting from the start-up of shear (waiting time $t_w=0$~s, symbols) along with fits to the data from MCT (lines)~\cite{Laurati_12}. The presence of a plateau in the quiescent MSD followed by an increase at long times signifies the caging of particles by their neighbours before diffusion at longer times takes place due to collective particle motion. In the steady state of shear, the MSDs exhibit faster dynamics than at rest, as seen by the earlier onset of diffusion and the correspondingly larger value of the
long-time diffusion coefficient. The dynamics become faster upon increasing $Pe_0$ and long-time diffusion begins earlier, suggesting that the shear rate determines the charateristic timescale of the dynamics. For the transient case with $t_w=0$~s, the MSDs follow the quiescent curve for low strains, whilst at long times displaying diffusive behaviour equivalent to the steady state curve, evidencing the shear-thinning nature of these dispersions. At intermediate times however, there is a rapid increase in the MSD with a growth exponent greater than the diffusive exponent of 1, i.e. super-diffusive motion is observed. The prominence of the super-diffusion is seen to increase with $Pe_0$, which is similar to the increase of the stress overshoot and is in agreement with predictions from MCT. The range of strains over which super-diffusion is displayed is essentially independent of $Pe_0$, being typically 1.5\%--6\%. That this range seems to be independent of shear rate and waiting time suggests the extent of the transient regime is related to a static lengthscale of the system such as average neighbour separation~\cite{Laurati_12}.

MCT relates the stress overshoot to the super diffusive regime of the MSD via the generalized shear modulus; in the quiescent state, the shear modulus is always positive, whilst under shear it exhibits a small negative portion which gives rise to both the stress overshoot and the super-diffusive regime of the MSD. The dip in the generalized shear modulus arises from the secondary $\alpha$-relaxation, which is attributed to the breaking up of nearest neighbour cages~\cite{Zausch_08}.

\subsection{Summary of transient properties under switch-on of shear}

A fuller picture of yielding can now be conjectured: at the start-up of shear, stresses build in the system due to the stretching of particle cages. The structural anisotropy reaches its maximum at strains similar to the maximum in the stress and corresponding to the point at which particle cages start to break up, essentially the yielding point. During cage yielding, super-diffusive dynamics are observed as particles move from their cages due to the shear. In the steady state, the system is essentially fluidlike, displaying diffusive dynamics and a less pronounced though still significant structural anisotropy induced by shear and due to the constant breaking and reformation of particle cages. In the limit of low shear rates, no stress overshoot is observed since stress is completely relaxed through Brownian diffusion before the cage is considerably distorted. At higher shear rates however, cage deformation is rapid enough that stress can be stored before yielding and relaxes partly at the steady state.

\section{Cessation of shear: switch off}
\label{sec:swoff}

\subsection{Rheology}

Upon sudden cessation of shear, the relaxation of dense colloidal suspensions displays behaviour that is heavily dependent on both the properties of the system, such as $\phi$, and on the previously sheared state~\cite{Zausch_09,Ballauff_13}. Through a combination of rheology on hard and soft spheres (systems as in \S\ref{sec:swon-rheo}), MD simulations and MCT, it was recently shown that previously sheared glasses can display history dependent residual stresses~\cite{Ballauff_13}. 

\begin{figure}
\begin{center}
\includegraphics[width=0.9\linewidth]{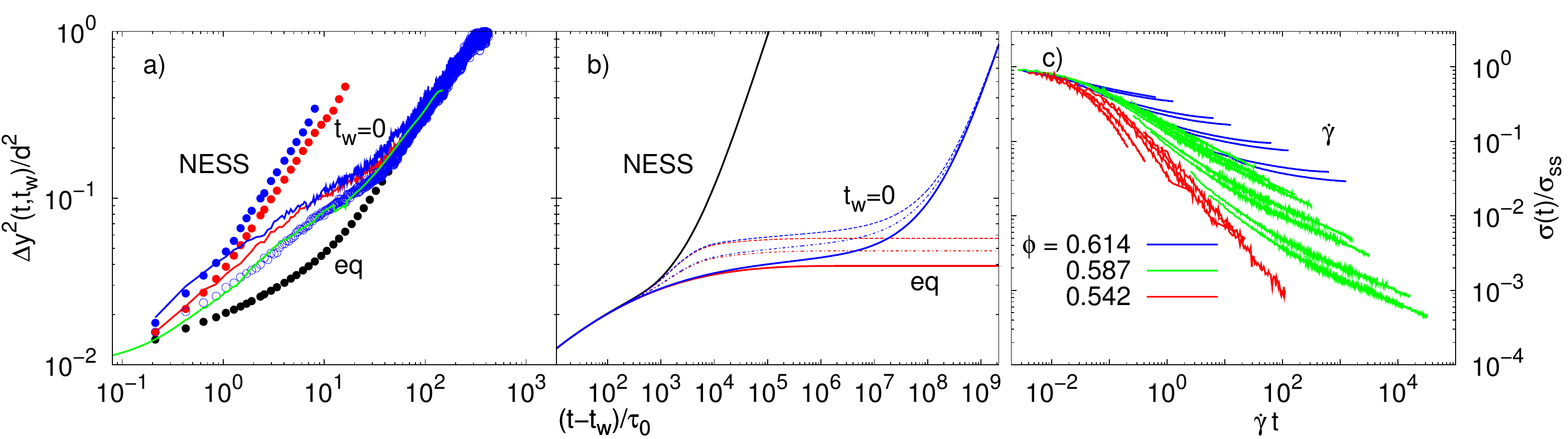}
\caption{(a) MSDs in the vorticity direction after cessation of shear from confocal microscopy on PMMA colloids at $\varphi=0.56$. The equilibrium (eq), sheared steady state (NESS) for $Pe_0=0.019$ (red) and $Pe_0=0.037$ (blue) and the transient data for $t_{\mathrm{w}}=0$~s (solid lines) and $t_{\mathrm{w}}=4\,\text{s}$ (open symbols) are shown along with a fit to the data from MCT-ITT (green line). (b) MSDs from schematic MCT-ITT model for two state points $\varepsilon=\pm10^{-4}$ (glass shown by red lines, liquid by blue lines) with $\dot\gamma t_w=0$ and $\dot\gamma\,t_w=0.01$, for $\dot\gamma\tau_0=10^{-5}$. (Labels as in (a).) (c) Stress relaxation after shear cessation on PMMA colloids at $\phi$ as indicated \cite{Ballauff_13}.\label{fig:sw-off}  }
\end{center}
\end{figure}

For dense liquids, the shear stress decays to zero on the structural-relaxation timescale, $\tau$, which is typically much larger than the timescale for free diffusion, $\tau_0$ (Fig.~\ref{fig:sw-off}(c)). However, as $\phi$ increases towards $\phi_{\mathrm{g}}$, $\tau$ grows beyond experimentally accessible times and a plateau in the decay of the shear stress develops. Experimentally measured stress plateaus decay very slowly, possibly as a result of creep, whereas within MCT, pre-sheared glassy states display a nonrelaxing persistent residual stress. This is due to the divergence of the structural relaxation time as full relaxation of density fluctuations is impeded by permanent local caging of particles~\cite{Ballauff_13}.

Whilst the yield stress of a concentrated colloidal system typically displays an increase with $Pe_0$, the residual stresses show a decrease, which suggests that the stress relaxation is more effective for strongly fluidised samples. Scaling the stresses by their flow-curve values, $\sigma_{\mathrm{ss}}(\dot{\gamma})$, and the time with $\dot{\gamma}$, a separation into two distinct classes of liquids and glasses is observed, Fig.~\ref{fig:sw-off}(c). The ability to rescale by $\dot{\gamma}$ is particularly remarkable given that the shear has already been stopped and the equations of motion contain no reference to the previous perturbation, i.e.~the relaxing state has some memory of the shear. The final stress decay for the observed liquids does not scale with $\dot{\gamma}$ as this is governed by the equilibrium relaxation time~\cite{Ballauff_13}.

\subsection{Microscopic Dynamics}

Similar to the case of start-up shear, further understanding can be gained by measuring the dynamics of the relaxing state by microscopy and simulations. The MSDs for dense fluid samples after cessation of shear as experimentally determined and calculated from ITT-MCT are shown in Fig.~\ref{fig:sw-off}(a,b). The short time dynamics follows the steady state curve, which is diffusive, for a time of $\dot{\gamma}t\approx0.1$, i.e.~even after cessation of shear the system shows shear-rate dependent dynamics. The transient regime displays sub-diffusion leading to long-time diffusive behaviour, resembling the dynamics of the quiescent state. This intermediate plateau is reminiscent of the quiescent case where it can be linked to particle caging. 

The evolution of the transient MSD for short waiting times can be predicted by ITT-MCT based upon the MSDs of the quiescent state and the steady sheared state (green line, Fig.~\ref{fig:sw-off}(a)). Within MCT, the intermediate plateau represents a second length scale beyond the quiescent localisation length, arising from the competition of shear-induced fluidisation and arrest after cessation, suggesting that this is linked to the residual stresses of the system. It is an effective stopping distance caused by the progression of the distribution function from the perturbed state to the quiescent state. 

For glassy states, the evolution of the dynamics after cessation of shear has been studied by MD simulations \cite{Ballauff_13}, but has yet to be measured experimentally due to limitations in the stability of the experimental setup. Moreover, highly concentrated samples can exhibit shear banding under low shear rates~\cite{Besseling_10}; the effect of this on the switch off structure can be investigated once sufficiently high shear rates are experimentally accessible. Overcoming these issues and studying both the dynamics and the structural origin of residual stresses after shear cessation will form the focus of future research.

\section{Large amplitude oscillatory shear (LAOS)}
\label{sec:laos}

Key aspects of the nonlinear rheology of viscoelastic dispersions and yielding soft solids is also displayed upon application of an oscillatory shear strain 
\begin{equation}
\gamma(t)=\gamma_0\sin(\omega t)\; ,
\label{strain} 
\end{equation} 
introducing the parameters strain amplitude $\gamma_0$ and {frequency $\omega$}. The response of a dispersion at long times, such that transient effects from switching on the strain have decayed, is dicussed~\cite{Siebenbuerger_12}.

\begin{figure}[t!]
\begin{center}
  \includegraphics[width=.45\linewidth]{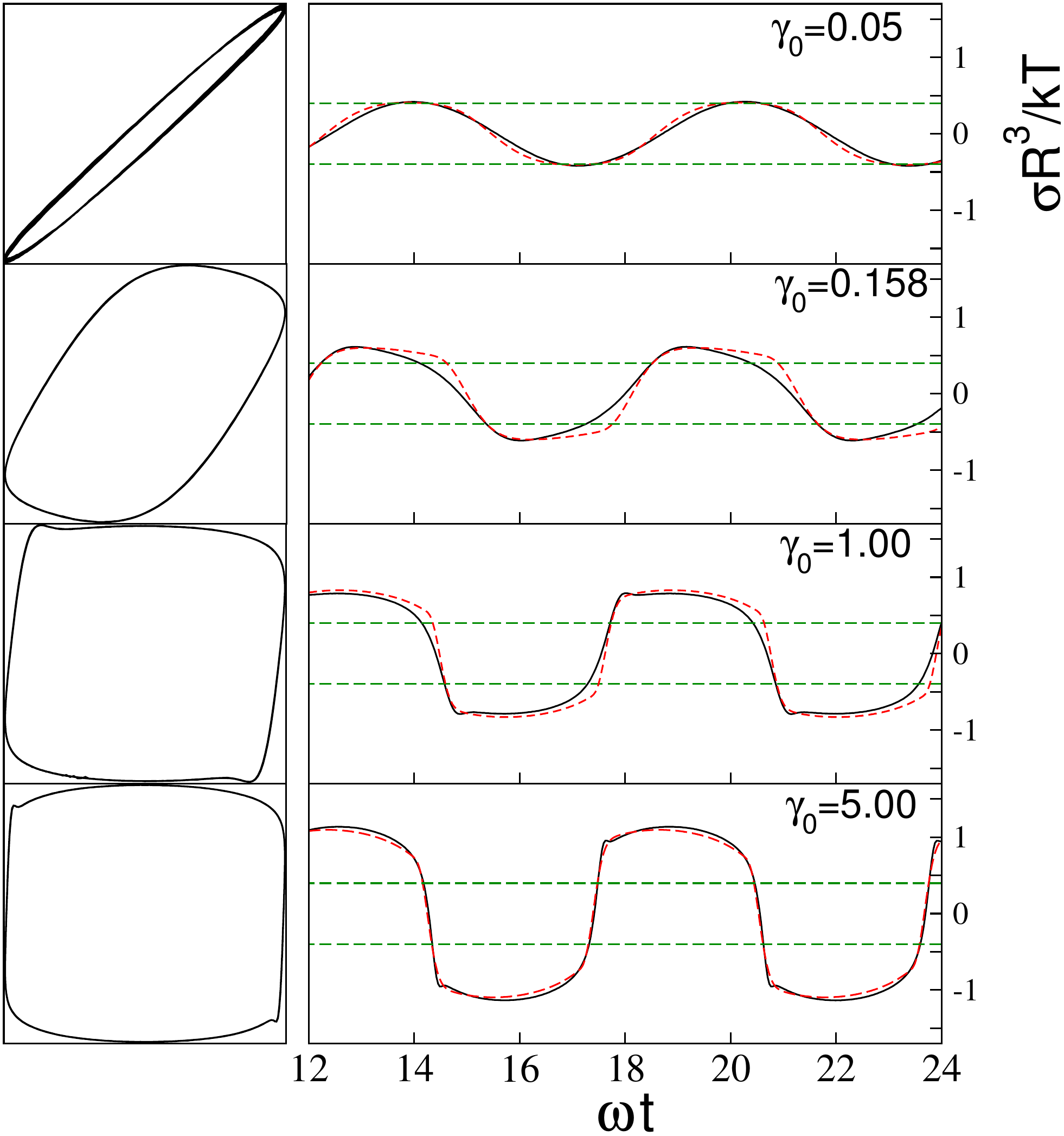}
\caption{Stress response measured in LAOS experiments for strain amplitudes $0.05 \le \gamma_0 \le 5$ (black lines) and associated Lissajous figures illustrating the nonlinear character of the response. The experiments on PNIPAM microgels were performed on a glassy sample at $T=15.1^{\circ}$C and a frequency of 1~Hz (corresponding to $Pe_{\omega}=0.02533$). The yield stress is indicated by the green dashed lines, theoretical results by red dashed lines \cite{brader2010a}.
\label{fig:laos}}
\end{center}
\end{figure}

The LAOS stress signals measured in PNIPAM microgel dispersions at a frequency in the {$\beta$-window} (0.01~Hz, corresponding to $Pe_{\omega}=6\pi\eta_s a^3\omega/k_BT=0.025$) are shown in the right-hand panels of Fig.~\ref{fig:laos}. The deformation varies from $\gamma_0=0.05$ in the linear regime with an almost entirely elastic response to $\gamma_0=5$ deep in the nonlinear regime with an almost purely viscous response. The increase in dissipation with increasing $\gamma_0$ is reflected in the increasing area enclosed by the Lissajous curves. For small strain amplitudes, the linear viscoelastic behaviour is indicated by the nearly perfect sinusoidal stress $\sigma(t)$. The time dependent signal becomes distorted away from a pure sinusoid when the peak of $\sigma(t)$ encounters the dynamic yield stress (green dashed lines, Fig.\ref{fig:laos}).
Theory (red dashed lines) predicts that for even lower frequencies, the peak maximum should approach the yield stress value, but this could not be tested experimentally. For the intermediate chosen frequency, the stress signal displays flattened asymmetric peaks at intermediate values of $\gamma_0$, consistent with a regime of cage breaking around $\gamma_0=\gamma_c$.  The schematic model calculations included in the figure are in good agreement with the experimental data. Data at higher frequencies show somewhat larger deviations from the model calculations. Similar information can be obtained from the `Lissajous curves' (left panels, Fig.~\ref{fig:laos}).

\section{Conclusions}

The transient response of dense colloidal suspensions both to the application of an external shear and to its cessation, have been studied with the aim of elucidating the shear-induced structural and dynamical changes in these systems. These insights have been made possible through advancements in mode coupling theory and experimental techniques, namely confocal microscopy and the associated data analysis.

Application of external shear results in cage deformations on the microscopic particle level as well as an increased (reduced) probability to find particles at contact in the compression (extension) direction, respectively, which reflects the build up of stresses. At the same time, the transient dynamics exhibit super-diffusive behaviour. Thus the applied shear enforces the decay of correlations even in a state that would otherwise be dynamically arrested. MCT links the super-diffusion to the overshoot observed in the rheological response. The external shear rate determines the relaxation rate, which is much faster than the internal structural relaxation rate, and allows for local particle motion leading to cage distortion and shear-thinning. With increasing strain and beyond the yield strain, a steady-state is reached that is characterised by a reduced though still significant structural anisotropy, purely diffusive dynamics and a fluid-like rheological response. After cessation of shear, the shear stress does not fully decay but history-dependent residual stresses remain, which are smaller in previously strongly fluidised samples. The dynamics are sub-diffusive with a plateau which represents a second length scale in addition to the quiescent localisation length. The localisation length is caused by the competition of shear-induced fluidisation and cessation-induced arrest, suggesting a link to the residual stress.

\newpage
We acknowledge support by the Deutsche Forschungsgemeinschaft (DFG) through the SFB-Transregio TR6 ``The Physics of Colloidal Dispersions in External Fields'' (project A6) during the period 2005--2013. We thank J.M.~Brader, M.~Kr\"uger, O.~Henrich, Th.~Voigtmann and F.~Weysser for joint work on this project. We also thank M.E.~Cates, N.~Koumakis, G.~Petekidis, M.~Siebenb\"urger, and M.~Ballauff for the fruitful cooperation and discussions, and J.~Horbach (project A5) for collaboration.

\bibliography{review_a6}

\begin{thebibliography}{100}

\bibitem{Pusey_91}
{\sc P.~N. Pusey},
\newblock {\em Liquids, Freezing and the Glass Transition, In: Proceedings of
  the Les Houches Summer School, J. P. Hansen, D. Levesque, and J. Zinn-Justin
  (eds.)},
\newblock Elsevier, Amsterdam, 1991.

\bibitem{Poon_00}
{\sc W.~C.~K. Poon},
\newblock {\em A day in the life of a hard-shpere suspension, In: Soft and
  Fragile Matter: Nonequilibrium Dynamics, Metastability and Flow, M. E. Cates
  and M. R. Evans (eds.)},
\newblock Institute of Physics Pub., copublished by Scottish Universities
  Summer School in Physics, Oxford, 2000.

\bibitem{McLeish_00}
{\sc T.~McLeish},
\newblock {\em Rheology of linear and branched polymers, In: Soft and Fragile
  Matter: Nonequilibrium Dynamics, Metastability and Flow, M. E. Cates and M.
  R. Evans (eds.)},
\newblock Institute of Physics Pub., copublished by Scottish Universities
  Summer School in Physics, Oxford, 2000.

\bibitem{brader2010b}
{\sc J.~M. Brader},
\newblock {\em J.~Phys.:~Condens.~Matter} {\bf 22}, 363101 (2010).

\bibitem{Hoover_68}
{\sc W.~G. Hoover} and {\sc F.~H. Ree},
\newblock {\em J. Chem. Phys.} {\bf 49}, 3609 (1968).

\bibitem{Pusey_86}
{\sc P.~N. Pusey} and {\sc {W. van Megen}},
\newblock {\em Nature} {\bf 320}, 340 (1986).

\bibitem{Gasser_09}
{\sc U.~Gasser},
\newblock {\em J. Phys.: Condens. Matter} {\bf 21}, 203101 (2009).

\bibitem{Pusey_09}
{\sc P.~N. Pusey}, {\sc E.~Zaccarelli}, {\sc C.~Valeriani}, {\sc E.~Sanz}, {\sc
  W.~C.~K. Poon}, and {\sc M.~E. Cates},
\newblock {\em Phil. Trans. R. Soc. A} {\bf 367}, 4993 (2009).

\bibitem{Woodcock_81}
{\sc L.~V. Woodcock},
\newblock {\em Ann. N. Y. Acad. Sci.} {\bf 37}, 274 (1981).

\bibitem{Pusey_87}
{\sc P.~N. Pusey} and {\sc {W. van Megen}},
\newblock {\em Phys. Rev. Lett.} {\bf 59}, 2083 (1987).

\bibitem{vanMegen_91}
{\sc {W. van Megen}} and {\sc P.~N. Pusey},
\newblock {\em Phys. Rev. A} {\bf 43}, 5429 (1991).

\bibitem{Goetze_92}
{\sc W.~G\"otze} and {\sc L.~Sj\"ogren},
\newblock {\em Rep. Prog. Phys.} {\bf 55}, 241 (1992).

\bibitem{Goetze}
{\sc W.~G\"otze},
\newblock {\em Complex Dynamics of Glass-Forming Liquids, A Mode-Coupling
  Theory},
\newblock Oxford University Press, 2009.

\bibitem{vanMegen_93}
{\sc {W. van Megen}} and {\sc S.~M. Underwood},
\newblock {\em Phys. Rev. Lett.} {\bf 70}, 2766 (1993).

\bibitem{vanMegen_94}
{\sc {W. van Megen}} and {\sc S.~M. Underwood},
\newblock {\em Phys. Rev. E} {\bf 49}, 4206 (1994).

\bibitem{Krieger_72}
{\sc I.~M. Krieger},
\newblock {\em Adv. Colloid Interface Sci.} {\bf 3}, 111 (1972).

\bibitem{Krieger_76}
{\sc I.~M. Krieger} and {\sc M.~Eguiluz},
\newblock {\em Trans. Soc. Rheol.} {\bf 20}, 29 (1976).

\bibitem{Ackerson_91}
{\sc B.~J. Ackerson},
\newblock {\em Physica A} {\bf 174}, 15 (1991).

\bibitem{Pham_06}
{\sc K.~N. Pham}, {\sc G.~Petekidis}, {\sc D.~Vlassopoulos}, {\sc S.~U.
  Egelhaaf}, {\sc P.~N. Pusey}, and {\sc W.~C.~K. Poon},
\newblock {\em Europhys. Lett.} {\bf 75}, 624 (2006).

\bibitem{Pham_08}
{\sc K.~N. Pham}, {\sc G.~Petekidis}, {\sc D.~Vlassopoulos}, {\sc S.~U.
  Egelhaaf}, {\sc W.~C.~K. Poon}, and {\sc P.~N. Pusey},
\newblock {\em J. Rheol.} {\bf 52}, 649 (2008).

\bibitem{Sentjabrskaja_13}
{\sc T.~Sentjabrskaja}, {\sc E.~Babaliari}, {\sc J.~Hendricks}, {\sc
  M.~Laurati}, {\sc G.~Petekidis}, and {\sc S.~U. Egelhaaf},
\newblock {\em Soft Matter} {\bf 9}, 4524 (2013).

\bibitem{Batchelor_77}
{\sc G.~K. Batchelor},
\newblock {\em J. Fluid Mech.} {\bf 83}, 97 (1977).

\bibitem{Petekidis_02A}
{\sc G.~Petekidis}, {\sc P.~N. Pusey}, {\sc A.~Moussa\"id}, {\sc S.~U.
  Egelhaaf}, and {\sc W.~C.~K. Poon},
\newblock {\em Phys. A} {\bf 306}, 334 (2002).

\bibitem{Petekidis_02}
{\sc G.~Petekidis}, {\sc A.~Moussa\"id}, and {\sc P.~N. Pusey},
\newblock {\em Phys. Rev. E} {\bf 66}, 051402 (2002).

\bibitem{Petekidis_03}
{\sc G.~Petekidis}, {\sc D.~Vlassopoulos}, and {\sc P.~N. Pusey},
\newblock {\em Faraday Discuss.} {\bf 123}, 287 (2003).

\bibitem{Pham_04}
{\sc K.~N. Pham}, {\sc S.~U. Egelhaaf}, {\sc A.~Moussa\"id}, and {\sc P.~N.
  Pusey},
\newblock {\em Rev. Sci. Instrum.} {\bf 75}, 2419 (2004).

\bibitem{Besseling_07}
{\sc R.~Besseling}, {\sc E.~R. Weeks}, {\sc A.~B. Schofield}, and {\sc W.~C.~K.
  Poon},
\newblock {\em Phys. Rev. Lett.} {\bf 99}, 028301 (2007).

\bibitem{Mutch_13}
{\sc K.~J. Mutch}, {\sc N.~Koumakis}, {\sc M.~Laurati}, and {\sc S.~U.
  Egelhaaf},
\newblock {\em to be published} .

\bibitem{DHaene_93}
{\sc P.~D'Haene}, {\sc J.~Mewis}, and {\sc G.~G. Fuller},
\newblock {\em J. Colloid Interface Sci.} {\bf 156}, 350 (1993).

\bibitem{Bender_96}
{\sc J.~W. Bender} and {\sc N.~J. Wagner},
\newblock {\em J. Rheol.} {\bf 40}, 899 (1996).

\bibitem{Maranzano_02}
{\sc B.~J. Maranzano} and {\sc N.~J. Wagner},
\newblock {\em J. Chem. Phys.} {\bf 117}, 10291 (2002).

\bibitem{Foss_00}
{\sc D.~R. Foss} and {\sc J.~F. Brady},
\newblock {\em J. Fluid Mech.} {\bf 407}, 167 (2000).

\bibitem{Brady_01}
{\sc J.~F. Brady},
\newblock {\em Chem. Eng. Sci.} {\bf 56}, 2921 (2001).

\bibitem{Nguyen_83}
{\sc Q.~D. Nguyen} and {\sc D.~V. Boger},
\newblock {\em J. Rheol.} {\bf 27}, 321 (1983).

\bibitem{Persello_94}
{\sc J.~Persello}, {\sc J.~Chang}, {\sc B.~Cabane}, {\sc A.~Magnin}, and {\sc
  J.~M. Piau},
\newblock {\em J. Rheol.} {\bf 38}, 1845 (1994).

\bibitem{Derec_03}
{\sc C.~Derec}, {\sc G.~Ducouret}, {\sc A.~Adjari}, and {\sc F.~Lequeux},
\newblock {\em Phys. Rev. E} {\bf 67}, 061403 (2003).

\bibitem{Mahaut_08}
{\sc F.~Mahaut}, {\sc X.~Chateau}, {\sc P.~Coussot}, and {\sc G.~Ovarlez},
\newblock {\em J. Rheol.} {\bf 52}, 287 (2008).

\bibitem{Khan_88}
{\sc S.~A. Khan}, {\sc C.~A. Schnepper}, and {\sc R.~C. Armstrong},
\newblock {\em J. Rheol.} {\bf 32}, 69 (1988).

\bibitem{Islam_04}
{\sc M.~T. Islam}, {\sc N.~Rodriguez-Hornedo}, {\sc S.~Ciotti}, and {\sc
  C.~Ackermann},
\newblock {\em Pharm. Res.} {\bf 21}, 1192 (2004).

\bibitem{Mohraz_05}
{\sc A.~Mohraz} and {\sc M.~J. Solomon},
\newblock {\em J. Rheol.} {\bf 49}, 657 (2005).

\bibitem{Becu_06}
{\sc L.~B\'ecu}, {\sc S.~Manneville}, and {\sc A.~Colin},
\newblock {\em Phys. Rev. Lett.} {\bf 96}, 138302 (2006).

\bibitem{Akcora_09}
{\sc P.~Akcora}, {\sc H.~Liu}, {\sc S.~K. Kumar}, {\sc J.~Moll}, {\sc Y.~Li},
  {\sc B.~C. Benicewicz}, {\sc L.~S. Schadler}, {\sc D.~Acehan}, {\sc A.~Z.
  Panagiotopoulos}, {\sc V.~Pryamitsyn}, {\sc V.~Ganesan}, {\sc J.~Ilavsky},
  {\sc P.~Thiyagarajan}, {\sc R.~H. Colby}, and {\sc J.~F. Douglas},
\newblock {\em Nature Mater.} {\bf 8}, 354 (2009).

\bibitem{Carrier_09}
{\sc V.~Carrier} and {\sc G.~Petekidis},
\newblock {\em J. Rheol.} {\bf 53}, 245 (2009).

\bibitem{Divoux_11}
{\sc T.~Divoux}, {\sc C.~Barentin}, and {\sc S.~Manneville},
\newblock {\em Soft Matter} {\bf 7}, 9335 (2011).

\bibitem{Berret_97}
{\sc J.-F. Berret},
\newblock {\em Langmuir} {\bf 13}, 2227 (1997).

\bibitem{Lerouge_98}
{\sc S.~Lerouge}, {\sc J.-P. Decruppe}, and {\sc C.~Humbert},
\newblock {\em Phys. Rev. Lett.} {\bf 81}, 5457 (1998).

\bibitem{Soltero_99}
{\sc J.~F.~A. Soltero}, {\sc F.~Bautista}, {\sc J.~E. Puig}, and {\sc
  O.~Manero},
\newblock {\em Langmuir} {\bf 15}, 1604 (1999).

\bibitem{Osaki_00}
{\sc K.~Osaki}, {\sc T.~Inoue}, and {\sc T.~Isomura},
\newblock {\em J. Polym. Sci. B} {\bf 38}, 2043 (200).

\bibitem{Islam_01}
{\sc M.~T. Islam} and {\sc L.~A. Archer},
\newblock {\em J. Polym. Sci. B} {\bf 39}, 2275 (2001).

\bibitem{Decruppe_01}
{\sc J.-P. Decruppe}, {\sc S.~Lerouge}, and {\sc J.-F. Berret},
\newblock {\em Phys. Rev. E} {\bf 63}, 022501 (2001).

\bibitem{Boukany_10}
{\sc P.~E. Boukany}, {\sc O.~Hemminger}, {\sc S.-Q. Wang}, and {\sc L.~J. Lee},
\newblock {\em Phys. Rev. Lett.} {\bf 105}, 027802 (2010).

\bibitem{Ganapathy_08}
{\sc R.~Ganapathy} and {\sc A.~K. Sood},
\newblock {\em J. Non-Newtonian Fluid Mech.} {\bf 149}, 78 (2008).

\bibitem{Ravindranath_08}
{\sc S.~Ravindranath} and {\sc S.-Q. Wang},
\newblock {\em J. Rheol.} {\bf 52}, 681 (2008).

\bibitem{Tapadia_04}
{\sc P.~Tapadia} and {\sc S.-Q. Wang},
\newblock {\em Macromolecules} {\bf 37}, 9083 (2004).

\bibitem{Tapadia_06}
{\sc P.~Tapadia} and {\sc S.-Q. Wang},
\newblock {\em Phys. Rev. Lett.} {\bf 96}, 016001 (2006).

\bibitem{larson}
{\sc R.~G. Larson},
\newblock {\em {The structure and rheology of complex fluids}},
\newblock Oxford University Press, New York, 1999.

\bibitem{Langer2011}
{\sc E.~Bouchbinder} and {\sc J.~S. Langer},
\newblock {\em Phys.~Rev.~Lett.} {\bf 106}, 148301 (2011).

\bibitem{Schall_07}
{\sc P.~Schall}, {\sc D.~A. Weitz}, and {\sc F.~Spaepen},
\newblock {\em Science} {\bf 318}, 1895 (2007).

\bibitem{SGR}
{\sc P.~Sollich}, {\sc F.~Lequeux}, {\sc P.~H{\'e}braud}, and {\sc M.~E.
  Cates},
\newblock {\em Phys. Rev. Lett.} {\bf 78}, 2020 (1997).

\bibitem{Siebenbuerger_12}
{\sc M.~Siebenb{\" u}rger}, {\sc M.~Fuchs}, and {\sc M.~Ballauff},
\newblock {\em Soft Matter} {\bf 8}, 4025 (2012).

\bibitem{Ikeda_12}
{\sc A.~Ikeda}, {\sc L.~Berthier}, and {\sc P.~Sollich},
\newblock {\em Phys. Rev. Lett.} {\bf 109}, 018301 (2012).

\bibitem{dhont}
{\sc J.~K.~G. Dhont},
\newblock {\em An Introduction to Dynamics of Colloids},
\newblock Elsevier Science, Amsterdam, 1996.

\bibitem{brady}
{\sc J.~Bergenholtz}, {\sc J.~Brady}, and {\sc M.~Vicic},
\newblock {\em J.~Fluid.~Mech.} {\bf 456}, 239 (2002).

\bibitem{Russel}
{\sc W.~Russel}, {\sc D.~A. Saville}, and {\sc W.~R. Schowalter},
\newblock {\em Colloidal dispersions},
\newblock Cambridge University Press, 1989.

\bibitem{Brady_93}
{\sc J.~F. Brady},
\newblock {\em J. Chem. Phys.} {\bf 99}, 567 (1993).

\bibitem{brader2011}
{\sc J.~M. Brader} and {\sc M.~Kr\"uger},
\newblock {\em Mol. Phys.} {\bf 109}, 1029 (2011).

\bibitem{Archer}
{\sc P.~Hopkins}, {\sc A.~Fortini}, {\sc A.~J. Archer}, and {\sc M.~Schmidt},
\newblock {\em J. Chem. Phys.} {\bf 133}, 224505 (2010).

\bibitem{vanMegen1998}
{\sc W.~van Megen}, {\sc T.~C. Mortensen}, {\sc S.~Williams}, and {\sc
  J.~M\"uller},
\newblock {\em Phys. Rev. E} {\bf 58}, 6073 (1998).

\bibitem{Kawasaki}
{\sc K.~Kawasaki} and {\sc J.~D. Gunton},
\newblock {\em Phys. Rev. A} {\bf 8}, 2048 (1973).

\bibitem{EvansITT}
{\sc D.~J. Evans} and {\sc G.~P. Morriss},
\newblock {\em Statistical Mechanics of Nonequilibrium Liquids},
\newblock Cambridge University Press, 2008.

\bibitem{brader2008}
{\sc J.~M. Brader}, {\sc M.~Cates}, and {\sc M.~Fuchs},
\newblock {\em Phys. Rev. Lett.} {\bf 101}, 138301 (2008).

\bibitem{brader2012}
{\sc J.~M. Brader}, {\sc M.~E. Cates}, and {\sc M.~Fuchs},
\newblock {\em Phys. Rev. E} {\bf 86}, 021403 (2012).

\bibitem{fuchs02}
{\sc M.~Fuchs} and {\sc M.~E. Cates},
\newblock {\em Phys.~Rev.~Lett.} {\bf 89}, 248304 (2002).

\bibitem{fuchs09}
{\sc M.~Fuchs} and {\sc M.~E. Cates},
\newblock {\em J.~Rheol.} {\bf 53}, 957 (2009).

\bibitem{brader2009}
{\sc J.~M. Brader}, {\sc {\relax Th}.~Voigtmann}, {\sc M.~Fuchs}, {\sc R.~G.
  Larson}, and {\sc M.~E. Cates},
\newblock {\em Proc.\ Natl.\ Acad.\ Sci.\ USA} {\bf 106}, 15186 (2009).

\bibitem{Amann_13}
{\sc C.~P. Amann}, {\sc M.~Siebenb\"urger}, {\sc M.~Kr\"uger}, {\sc
  F.~Weysser}, {\sc M.~Ballauff}, and {\sc M.~Fuchs},
\newblock {\em J. Rheol.} {\bf 57}, 149 (2013).

\bibitem{Laurati_12}
{\sc M.~Laurati}, {\sc K.~J. Mutch}, {\sc N.~Koumakis}, {\sc J.~Zausch}, {\sc
  C.~P. Amann}, {\sc A.~B. Schofield}, {\sc G.~Petekidis}, {\sc J.~F. Brady},
  {\sc J.~Horbach}, {\sc M.~Fuchs}, and {\sc S.~U. Egelhaaf},
\newblock {\em J. Phys.: Condens. Matter} {\bf 24}, 464104 (2012).

\bibitem{Koumakis_12a}
{\sc N.~Koumakis}, {\sc A.~Pamvouxoglou}, {\sc A.~S. Poulos}, and {\sc
  G.~Petekidis},
\newblock {\em Soft Matter} {\bf 8}, 4271 (2012).

\bibitem{Crassous_08}
{\sc J.~J. Crassous}, {\sc M.~Siebenb\"urger}, {\sc M.~Ballauff}, {\sc
  M.~Drechsler}, {\sc D.~Hajnal}, {\sc O.~Henrich}, and {\sc M.~Fuchs},
\newblock {\em J. Chem. Phys.} {\bf 128}, 204902 (2008).

\bibitem{Siebenbuerger_09}
{\sc M.~Siebenb{\" u}rger}, {\sc M.~Fuchs}, {\sc H.~Winter}, and {\sc
  M.~Ballauff},
\newblock {\em J. Rheol.} {\bf 53}, 707 (2009).

\bibitem{Varnik_04}
{\sc F.~Varnik}, {\sc L.~Bocquet}, and {\sc J.-L. Barrat},
\newblock {\em J. Chem. Phys.} {\bf 120}, 2788 (2004).

\bibitem{Zausch_08}
{\sc J.~Zausch}, {\sc J.~Horbach}, {\sc M.~Laurati}, {\sc S.~U. Egelhaaf}, {\sc
  J.~M. Brader}, {\sc T.~Voigtmann}, and {\sc M.~Fuchs},
\newblock {\em J. Phys.: Condens. Matter} {\bf 20}, 404210 (2008).

\bibitem{Koumakis_12}
{\sc N.~Koumakis}, {\sc M.~Laurati}, {\sc S.~U. Egelhaaf}, {\sc J.~F. Brady},
  and {\sc G.~Petekidis},
\newblock {\em Phys. Rev. Lett.} {\bf 108}, 098303 (2012).

\bibitem{vanBlaaderen_95}
{\sc A.~{van Blaaderen}} and {\sc P.~Wiltzius},
\newblock {\em Science} {\bf 270}, 1177 (1995).

\bibitem{Weeks_99}
{\sc E.~R. Weeks}, {\sc J.~C. Crocker}, {\sc A.~C. Levitt}, {\sc A.~Schofield},
  and {\sc D.~A. Weitz},
\newblock {\em Science} {\bf 287}, 627 (1999).

\bibitem{Dinsmore_01}
{\sc A.~D. Dinsmore}, {\sc E.~R. Weeks}, {\sc V.~Prasad}, {\sc A.~C. Levitt},
  and {\sc D.~A. Weitz},
\newblock {\em Appl. Opt.} {\bf 40}, 4152 (2001).

\bibitem{Derks_04}
{\sc D.~Derks}, {\sc H.~Wisman}, {\sc A.~{van Blaaderen}}, and {\sc A.~Imhof},
\newblock {\em J. Phys.: Condens. Matter} {\bf 16}, S3917 (2004).

\bibitem{Smith_07}
{\sc P.~A. Smith}, {\sc G.~Petekidis}, {\sc S.~U. Egelhaaf}, and {\sc W.~C.~K.
  Poon},
\newblock {\em Phys. Rev. E} {\bf 76}, 041402 (2007).

\bibitem{Chen_10}
{\sc D.~Chen}, {\sc D.~Semwogerere}, {\sc J.~Sato}, {\sc V.~Breedveld}, and
  {\sc E.~R. Weeks},
\newblock {\em Phys. Rev. E} {\bf 81}, 011403 (2010).

\bibitem{Cheng_11}
{\sc X.~Cheng}, {\sc J.~H. McCoy}, {\sc J.~N. Israelachvili}, and {\sc
  I.~Cohen},
\newblock {\em Science} {\bf 333}, 1276 (2011).

\bibitem{Ballesta_08}
{\sc P.~Ballesta}, {\sc R.~Besseling}, {\sc L.~Isa}, {\sc G.~Petekidis}, and
  {\sc W.~C.~K. Poon},
\newblock {\em Phys. Rev. Lett.} {\bf 101}, 258301 (2008).

\bibitem{Besseling_09}
{\sc R.~Besseling}, {\sc L.~Isa}, {\sc E.~R. Weeks}, and {\sc W.~C.~K. Poon},
\newblock {\em Adv. Colloid Interface Sci.} {\bf 146}, 1 (2009).

\bibitem{Besseling_10}
{\sc R.~Besseling}, {\sc L.~Isa}, {\sc P.~Ballesta}, {\sc G.~Petekidis}, {\sc
  M.~E. Cates}, and {\sc W.~C.~K. Poon},
\newblock {\em Phys. Rev. Lett.} {\bf 105}, 268301 (2010).

\bibitem{Dutta_13}
{\sc S.~K. Dutta}, {\sc A.~Mbi}, {\sc R.~C. Arevalo}, and {\sc D.~L. Blair},
\newblock {\em Rev. Sci. Instrum.} {\bf 84}, 063702 (2013).

\bibitem{Crocker_96}
{\sc J.~C. Crocker} and {\sc D.~G. Grier},
\newblock {\em J. Colloid Interface Sci.} {\bf 179}, 298 (1996).

\bibitem{Jenkins_08}
{\sc M.~C. Jenkins} and {\sc S.~U. Egelhaaf},
\newblock {\em Adv. Colloid Interface Sci.} {\bf 136}, 65 (2008).

\bibitem{Semwogerere_05}
{\sc D.~Semwogerere} and {\sc E.~R. Weeks},
\newblock {\em Confocal microscopy, In: Encyclopedia of Biomaterials and
  Biomedical Engineering, G. Wnek and G. Bowlin (eds.)},
\newblock Taylor \& Francis, 2005.

\bibitem{Koumakis_long}
{\sc N.~Koumakis}, {\sc M.~Laurati}, {\sc K.~J. Mutch}, {\sc S.~U. Egelhaaf},
  {\sc J.~F. Brady}, and {\sc G.~Petekidis},
\newblock {\em to be published} .

\bibitem{Zausch_09}
{\sc J.~Zausch} and {\sc J.~Horbach},
\newblock {\em Europhys. Lett.} {\bf 88}, 60001 (2009).

\bibitem{Ballauff_13}
{\sc M.~Ballauff}, {\sc J.~M. Brader}, {\sc S.~U. Egelhaaf}, {\sc M.~Fuchs},
  {\sc J.~Horbach}, {\sc N.~Koumakis}, {\sc M.~Kr\"uger}, {\sc M.~Laurati},
  {\sc K.~J. Mutch}, {\sc G.~Petekidis}, {\sc M.~Siebenb\"urger}, {\sc
  T.~Voigtmann}, and {\sc J.~Zausch},
\newblock {\em Phys. Rev. Lett.} {\bf 110}, 215701 (2013).

\bibitem{brader2010a}
{\sc J.~M. Brader}, {\sc M.~Siebenb\"urger}, {\sc M.~Ballauff}, {\sc
  K.~Reinheimer}, {\sc M.~Wilhelm}, {\sc S.~J. Frey}, {\sc F.~Weysser}, and
  {\sc M.~Fuchs},
\newblock {\em Phys.~Rev.~E} {\bf 82}, 061401 (2010).

\end{thebibliography}

\end{document}